# Computational Exploration of Magnetic Saturation and Anisotropy Energy for Nonstoichiometric Ferrite Compositions

Venkata Rohit Punyapu[1,2], Jiazhou Zhu[1,3,Δ], Paul Meza-Morales[1,4,Δ], Anish Chaluvadi[1,5,6], O. Thompson Mefford[5,7], and Rachel B. Getman[1,2,*]

[1] Department of Chemical & Biomolecular Engineering, Clemson University, Clemson, SC 29634, USA
[2] Current affiliation: William G. Lowrie Department of Chemical and Biomolecular Engineering, The Ohio State University, Columbus, OH 43210, USA
[3] Current affiliation: Suzhou Novartis Technical Development Co., Ltd, Changshu, China
[4] Current affiliation: Intel Corporation, Hillsboro, Oregon, 97124, USA
[5] Department of Materials Science and Engineering, Clemson University, Clemson, SC 29634, USA
[6] Current affiliation: Department of Materials Science and Metallurgy, University of Cambridge, Cambridge, Cambridgeshire CB3 0FS, UK
[7] Department of Bioengineering, Clemson University, Clemson, SC 29634, USA
[Δ] Equal authorship
[*] Corresponding author:  getman.11@osu.edu

**Abstract.** A grand challenge in materials research is identifying the relationship between composition and performance. Herein, we explore this relationship for magnetic properties, specifically magnetic saturation ($M_s$) and magnetic anisotropy energy (MAE) of ferrites. Ferrites are materials derived from magnetite (which has the chemical formulae $Fe_3O_4$) that comprise metallic elements in some combination such as Fe, Mn, Ni, Co, Cu and Zn. They are used in a variety of applications such as electromagnetism, magnetic hyperthermia, and magnetic imaging. Experimentally, synthesis and characterization of magnetic materials is time consuming. In order to create insight to help guide synthesis, we compute the relationship between ferrite composition and magnetic properties using density functional theory (DFT). Specifically, we compute $M_s$ and MAE for 571 ferrite structures with the formulae $M1_xM2_yFe_{3-x-y}O_4$, where M1 and M2 can be Mn, Ni, Co, Cu and/or Zn and $0 \leq x \leq 1$ and $y = 1 - x$. By varying composition, we were able to vary calculated values of $M_s$ and MAE by up to $9.6 \times 10^5$ A m$^{-1}$ and $14.08 \times 10^5$ J m$^{-3}$, respectively. Our results suggest that composition can be used to optimize magnetic properties for applications in heating, imaging, and recording. This is mainly achieved by varying $M_s$, as these applications are more sensitive to variation in $M_s$ than MAE.

**1.0 Introduction.** The magnetite-derived ferrites composed of metallic elements in some combination such as Fe, Mn, Ni, Co, Cu and Zn have been widely studied for their structure and magnetic properties [1–6]. Typical ferrites have spinel-type (normal, inverse) crystal structures with $O^{2-}$ anions packed in a face-centered cubic (fcc) arrangement, such that there are two types of sites between them, i.e., tetrahedrally and octahedrally coordinated sites (see Figure 1). The general empirical formula for the stoichiometric class of ferrites is $M_xFe_{3-x}O_4$, where M can be different substituent metals (e.g., Mn, Ni, Co, Cu, Zn and other divalent metal cations) and $0 \leq x \leq 3$. Non-stoichiometric ferrites, i.e., materials with the general formula $M1_xM2_yFe_{3-x-y}O_4$, where M1 and M2 can be Mn, Ni, Co, Cu and/or Zn and $0 \leq x \leq 1$ and $y = 1 - x$, offer even greater compositional diversity.

Ferrite nanoparticles are widely used in the cores of transformers, antenna rods, electromagnets, and magnets used in imaging applications [7–11]. Another well-studied application is magnetically mediated energy delivery, which has most often been applied to biomedical devices (e.g., heating a cell via magnetic hyperthermia, i.e., MagMED) and more recently been applied to catalysts (e.g., supplying the heat needed to break and form chemical bonds via magnetic induction heating; i.e., MIH) [12–16]. Specifically, an oscillating magnetic field is applied, and hysteresis in the magnetic properties of

_________________________________________________________________________________

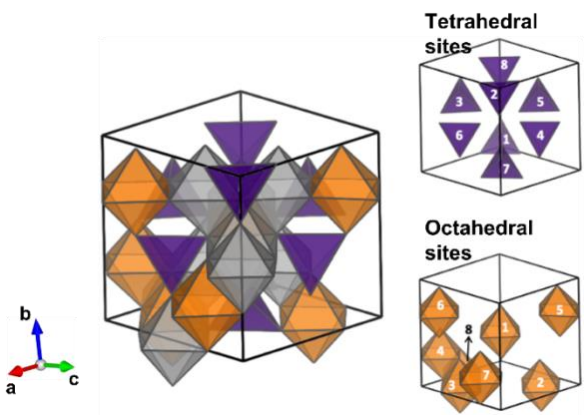

**Figure 1.** Left: Polyhedral representation of a bulk ferrite structure with stoichiometry $Fe_{24}O_{32}$. Purple = tetrahedral sites, orange and gray = octahedral sites. Right: Sites where substitutions were considered in this work.

the nanoparticle during oscillation results in conversion of magnetic energy into thermal energy [8,11–13]. The heat generated in hysteretic losses is due to Nèel and Brown relaxations, which are determined by the magnetic saturation ($M_s$), magnetic anisotropy energy (MAE), and the size of the nanoparticle [17,18]. $M_s$ and MAE, in turn, are determined by the particle composition [19–22].

A benefit to ferrites is that the compositions can be tuned. Indeed, multiple groups have investigated the influence of ferrite composition on performance for magnetic hyperthermia [17,23]. Specifically, these groups showed how varying composition results in values of $M_s$ and MAE that vary by up to a full order of magnitude. They further estimated the potential for heat generation as a function of size and composition, showing that this value could be varied by two orders of magnitude. Composition is important in other applications as well. For example, Co, Ni, Fe and Cu ferrites perform well for catalysis due to large hysteresis losses [12–14,16,21,24], while Zn, Co and Mn ferrites are often used in magnetic resonance imaging (MRI) because of the resulting high $M_s$ from their substitution [4,11,25]. An understanding of how composition influences magnetic properties is hence imperative to maximizing performance for the variety of applications that utilize ferrites and other magnetic materials.

While $M_s$ and MAE as well as other magnetic properties can be measured experimentally, the process is laborious; often requiring multiple attempts at synthesis to achieve the expected structure in addition to state-of-the-art measurement techniques to learn the magnetic properties [26–28]. Further, general rules linking magnetic properties to composition do not yet exist [1,17,20,25,29,30]. Filling this knowledge gap would greatly reduce the time and money required to design magnetic materials and devices for a wide range of applications; however, it would be impossible to accomplish this with experiments alone. On the other hand, computational approaches can provide estimates of magnetic properties relatively quickly. In such approaches, magnetic properties of model structures are computed with quantum mechanics. While these model structures are simplifications of the structures used in real-life applications – which is required for computational feasibility – they provide useful estimates of the magnetic properties of a given composition and are vastly more efficient at doing so than experiments. A database of magnetic material compositions and their associated magnetic properties would greatly facilitate design of magnetic materials for a variety of applications.

To this end, in this work we generate a database of ferrite compositions and their associated magnetic properties. Specifically, we compute values of $M_s$ and MAE for singly and doubly substituted non-stoichiometric ferrites using density functional theory (DFT). We investigate 571 total compositions and create an open access database that includes each composition's specific crystal structure (either normal or inverse spinel), calculated $M_s$, and calculated MAE. We further provide insight about the influence of composition on $M_s$ and MAE, showing that these values can vary by up to $10^3$ A m$^{-1}$ and $10^6$ J m$^{-3}$, respectively.

**2.0 Methodology**

**2.1 Ferrite Model Setup.** The ferrite model employed herein is based on the calculated bulk unit cell of magnetite ($Fe_3O_4$). We specifically employ a unit cell with space group of $F\bar{d}3m$. To create models with diverse compositions, we use eight repeats of the formula unit, giving a base stoichiometry of $Fe_{24}O_{32}$ (Figure 1). This model comprises eight Fe ions in tetrahedral sites (purple tetrahedra in Figure 1) and

___

**Table 1.** General compositions considered in this work, along with their range of $M_s$, range of MAE, and most prominent crystal structure.

| General Composition | Total Structures | $M_s \times 10^5$ [A m$^{-1}$] | MAE $\times 10^5$ [J m$^{-3}$] | Crystal Structure |
|---|---|---|---|---|
| **FeCu** | 24 | 1.2 – 9.4 | 1.3 – 5.5 | 62.5% Normal spinel |
| **FeMn** | 75 | 2.0 – 5.3 | 0.1 – 6.4 | 62.5% Inverse spinel |
| **FeCoCu** | 69 | 1.2 – 9.6 | 0.6 – 11.4 | 61.2% Inverse spinel |
| **FeMnCo** | 187 | 3.3 – 7.4 | 0.07 – 8.5 | 66.6% Normal spinel |
| **FeNiZn** | 34 | 0.3 – 8.9 * | 0.05 – 3.7 | 93.5% Inverse spinel |
| **FeMnNi** | 28 | 2.5 – 7.5 | 0.1 – 6.8 | 76.1% Inverse spinel |
| **FeNiCo** | 47 | 0.6 – 4.8 | 0.02 – 14.1 | 93.5% Inverse spinel |
| **FeNiCu** | 78 | 1.2 – 7.5 | 0.1 – 4.3 | 92% Inverse spinel |
| **FeCoZn** | 29 | 0.4 – 8.8 * | 0.06 – 11.1 | 64.5% Inverse spinel |

*Not included here is the $M_s$ of $Zn_8Fe_{16}O_{32}$ which has a $M_s$ of 0.04 A m$^{-1}$.

sixteen Fe ions in octahedral sites (orange and gray octahedra in Figure 1). Up to eight Fe ions are substituted with Mn, Ni, Co, Cu, and/or Zn in the purple and orange sites labeled 1 through 8 in Figure 1. We consider both singly (e.g., $Cu_1Fe_{23}O_{32}$, $Mn_7Fe_{17}O_{32}$) and doubly substituted (e.g., $Mn_2Co_2Fe_{20}O_{32}$, $Ni_1Cu_2Fe_{21}O_{32}$) ferrites in this work. Singly substituted ferrites include substitutions involving Mn and Cu and are denoted FeMn and FeCu for simplicity. Similarly, doubly substituted ferrites include combinations of Co and Cu, Ni and Zn, Co and Ni, Mn and Ni, Cu and Ni, Mn and Co, and Co and Zn, and are denoted FeCoCu, FeNiZn, FeCoNi, FeMnNi, FeCuNi, FeMnCo, and FeCoZn, respectively. The number of each general composition considered in this work is provided in Table 1. In total, 571 structures are considered, of which 99 are singly substituted and 472 are doubly substituted. Prior literature suggests that substitution into the same type of site (i.e., either tetrahedral or octahedral) but different location within the crystal lattice (e.g., different numbered tetrahedron or octahedron in Figure 1) has an influence on magnetic properties [22]. Hence, we also consider structures that have the same composition and substitution into the same *type* of site, but with the metal ions substituted into different tetrahedra or octahedra (e.g., $Cu_1^{tet,site1}Fe_{23}O_{32}$ and $Cu_1^{tet,site4}Fe_{23}O_{32}$). In this way, out of 571 structures, there are 204 unique compositions.

In general, ferrites can crystalize in either the normal (e.g., $[M1_x^{2+}M2_y^{2+}]^{tet}[Fe^{3+}]^{oct}_2O_4$) or inverse spinel structure (e.g., $[Fe^{3+}]^{tet}[M1_x^{2+}M2_y^{}Fe^{3+}]^{oct}_2O_4$) [5,6]. We consider both structures for each composition. Magnetic properties are reported for the structure that gives the lowest electronic energy in DFT. In rare cases, the electronic structure of one structure (i.e., either normal or inverse spinel) did not converge. In these cases, magnetic properties are reported for the structure that converged. The specific models employed in this work are available in the ioChem-BD database [31] along with their calculated $M_s$ and MAE.

**2.2 Magnetic property calculations.** $M_s$ is computed as

$$M_s = \frac{\text{Total magnetic moment} \times \mu_B}{\text{unit cell volume}} \quad \text{Eq. (1)}$$

where the total magnetic moment is the total number of unpaired electrons in the unit cell calculated in DFT, and $\mu_B$ is the Bohr magneton equal to $9.27 \times 10^{-24}$ A m$^2$. MAE is calculated as the difference in energy between the hard axis and the easy axis [32], i.e.,

$$\text{MAE} = E_\text{hard} - E_\text{easy} \quad \text{Eq. (2)}$$

where $E$ is the electronic energy calculated in DFT. The [0,0,1], [1,0,0] and [0,1,0] crystallographic directions are evaluated as the easy and hard axes for each composition. These directions were chosen since test calculations on the [0,1,1], [1,0,1], [1,1,0] and [1,1,1] directions often resulted in electronic energies significantly more positive than the

___

[0,0,1], [1,0,0] and [0,1,0] directions, suggesting that the [0,0,1], [1,0,0] and [0,1,0] directions are more reliable for large-scale DFT calculations. Among these three directions, the direction that gave the lowest electronic energy was taken as the easy axis, and the direction that resulted in the highest electronic energy was taken as the hard axis. Calculated easy and hard axis directions for singly substituted ferrites partially agree with prior results. For instance, in $Co_8Fe_{16}O_{32}$, the easy axis was found to be [100] in agreement with our results [33], while in $Ni_8Fe_{16}O_{32}$, the easy axis was found to be [111] [34], which does not match with our results (which determined the easy axis to be [010]). However, we find that MAEs calculated in this work generally follow experimental trends in cases where such data is available experimentally. Further details are provided in SI Section S7.

**2.3 Data Visualization.** The values of $M_s$ and MAE as functions of composition for doubly substituted ferrites are presented as contour plots. These plots interpolate between explicitly calculated points in order to create a continuous colormap. Each colormap is based on 18 – 31 explicitly calculated data points (see SI Section S6 for a sample plot with only data points). This is done using the OriginPro software [35] using the data boundary algorithm without smoothening. A total of 27 plots are generated for FeCoCu, FeNiZn, FeCoNi, FeMnNi, FeCuNi, FeMnCo FeCoZn, FeCu and FeMn, i.e., 9 for crystal structure, 9 for $M_s$ and 9 for MAE.

**2.4 Density Functional Theory Calculations.**
DFT calculations are performed using the Vienna Ab initio Simulation Package (VASP) [36,37]. Both collinear (i.e., all spins are aligned along the [0,0,1] direction) and non-collinear (i.e., spins are aligned along the [1,0,0], [0,1,0] and [0,0,1] directions) calculations are performed. Non-collinear calculations are always started from the wavefunction and charge density generated from a collinear calculation on the same system. Initial guesses for magnetic moments of the Fe, Mn, Co, Ni, Cu, and Zn cations are based on experimental findings by de Berg et al. [38] and reported in SI Section S9. The DFT+U formalism [30,39–41] is employed to capture the strong Coulombic repulsion on 3d electrons and to prevent the delocalization of electrons in these semiconducting materials. We specifically employ an effective U parameter, $U_{eff}$, equal to U − J, where U and J are the spherically averaged screened Coulomb and Exchange energies, respectively [42]. Values of U and J used to compute crystal structure and $M_s$ are taken from the Materials Project Database [43]. These values are provided in SI Section S9. Calculation of MAE requires a more stringent value of J in order to capture the spin orbit interaction [30] and achieve the magnetic ground state energies. We hence varied this value while holding values of U constant at the values taken from the Materials Project Database [43] and compared the resulting MAE with values from experiment [7,44] (see SI Section S10). These calculations were specifically done for the stoichiometric ferrites, i.e., $Fe_{24}O_{32}$, $Mn_8Fe_{16}O_{32}$, $Co_8Fe_{16}O_{32}$, $Ni_8Fe_{16}O_{32}$, $Cu_8Fe_{16}O_{32}$, and $Zn_8Fe_{16}O_{32}$. Resulting values of J were then used for the corresponding metal cations when calculating MAE values for the non-stoichiometric ferrites. We found that a value of $U_{eff}$ of 1.5 eV resulted in values of MAE that were in good agreement with experiment for the stoichiometric ferrites. Hence, J values for MAE calculations were taken as 1.5 eV – U.

In all calculations, electron exchange and correlation are treated using the Perdew-Burke-Ernzerhof (PBE) form of the generalized gradient approximation [45], the energies of core electrons are simulated with projector augmented wave (PAW) pseudopotentials [37,46] up to a cut-off energy of 550 eV, and spin polarization is turned on. Gamma-centered k-point meshes of 4×4×4 are used to sample the first Brillouin zones. Validation of the cut-off energy and k-point mesh are provided in SI Section S10. Electronic structures are calculated self-consistently and considered to be converged when the difference in energy between subsequent iterations falls below $10^{-6}$ eV for MAE calculations and $10^{-5}$ eV for everything else. During geometry relaxations, all atom positions as well as the unit cell shape and volume are allowed to relax. This strategy is validated in SI Section S10. Unit cell relaxations are considered to be converged when the absolute values of the forces on all atoms are smaller than 0.02 eV/Å. Most unit cells become slightly non-cubic during relaxation; however, the deviation from cubic is typically less than 0.3°. Examples of VASP input files are available in SI Section S9.

___

## 3.0 Results.

**3.1 Crystal Structure.** Calculated crystal structure preferences are shown in Figures 2 and S1 and tabulated in Table 1. We find that the stoichiometric ferrites $Fe_{24}O_{32}$, $Co_8Fe_{16}O_{32}$, $Ni_8Fe_{16}O_{32}$, $Cu_8Fe_{16}O_{32}$, and $Zn_8Fe_{16}O_{32}$ crystallize in the inverse spinel structure, whereas $Mn_8Fe_{16}O_{32}$ crystallizes in the normal spinel structure, in agreement with experiments [44,47–53]. In general, we find that non-stoichiometric ferrites largely crystallize in the inverse spinel structure (see Table 1). The exceptions are FeCu (Figure S1i) and FeMnCo (Figure 2f) which form mostly normal spinel structures (FeCu at low Cu content and FeMnCo at high Mn content). Of the remaining compositions that we investigated, FeNiCo forms inverse spinel structures over the majority of compositional space (Figure 2a) [52]. FeNiCu and FeNiZn also form mostly inverse spinel structures (Figure S1a and S1d). The remaining compositions form a mixture of normal and inverse spinel structures, depending on the composition (see Table 1 and Figures S1b, c and e). For example, FeMnNi forms a normal spinel structure at high Mn content and inverse spinel otherwise (Figure S1c), while FeCoZn and FeCoCu form inverse spinel structures at high Zn and Cu content and a mixture of inverse and normal spinel otherwise (Figure S1b and S1e).

**3.2 Saturation Magnetization.** Calculated $M_s$ are shown in Figures 3 and S2 and tabulated in Table 1. $M_s$ for the stoichiometric ferrites $Fe_{24}O_{32}$, $Mn_8Fe_{16}O_{32}$, $Co_8Fe_{16}O_{32}$, $Ni_8Fe_{16}O_{32}$, $Cu_8Fe_{16}O_{32}$, and $Zn_8Fe_{16}O_{32}$ are 4.8, 4.9, 3.6, 2.5, and 1.2, and $0.4 \times 10^{-6} \times 10^5$ A m$^{-1}$, respectively, approximately following the trends of the substituent cations themselves determined experimentally [17,54]. $M_s$ for most of the non-stoichiometric ferrites show large variations with composition, spanning more than $5 \times 10^5$ A m$^{-1}$. Exceptions to this are compositions involving Mn, which have narrower ranges of $M_s$ and hence cannot be as finely tuned as the other compositions investigated in this work. Comparing Figures S1 and S2, we observe the compositions that crystallize in the normal spinel structure exhibit higher $M_s$ than compositions that crystallize in the inverse spinel structure. For example, values of $M_s$ in the largely inverse spinel regions of FeNiZn, FeCoCu, FeCoZn and FeMnNi are relatively low, whereas values of $M_s$ in the

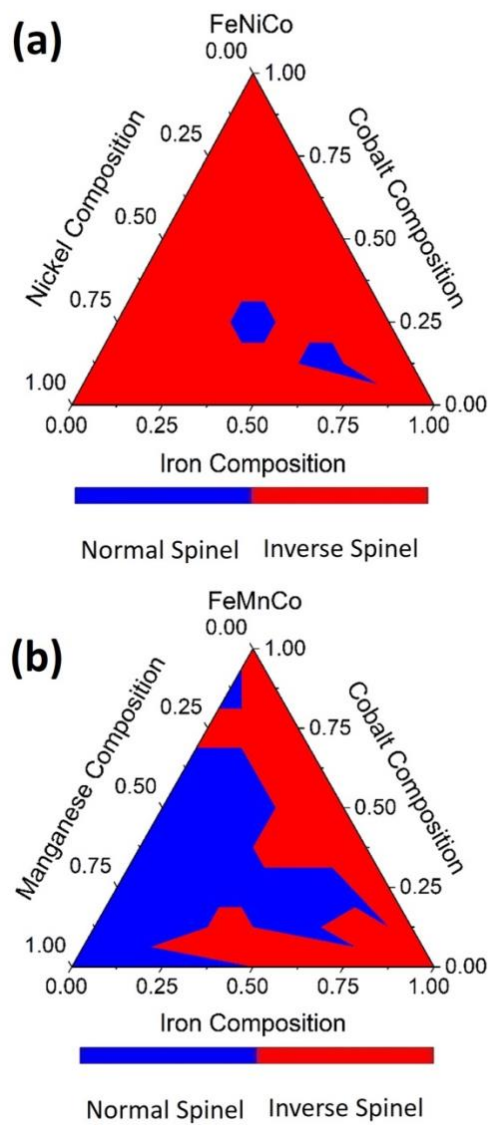

**Figure 2.** Calculated crystal structures of FeNiCo (a) and FeMnCo (b). Fe and substituent compositions span from 0 to 16 and 0 to 8, respectively, with the stoichiometric ferrites represented at the vertices.

normal spinel regions of these compositions are higher. In fact, the normal spinel regions of FeNiZn, FeCoCu and FeCoZn exhibit some of the highest $M_s$ calculated in this work, with compositions such as $Co_4Cu_1Fe_{19}O_{32}$ and $Co_5Zn_1Fe_{18}O_{32}$ exhibiting $M_s$ of 9.6 and $8.8 \times 10^5$ A m$^{-1}$, respectively. Conversely, the inverse spinel regions of these compositions exhibit some of the lowest values of $M_s$ calculated in this work, with compositions such as $Co_1Zn_7Fe_{16}O_{32}$ and $Co_6Zn_2Fe_{16}O_{32}$ (Figure 3b) exhibiting $M_s$ of 0.45 and $0.74 \times 10^5$ A m$^{-1}$ respectively. These

___

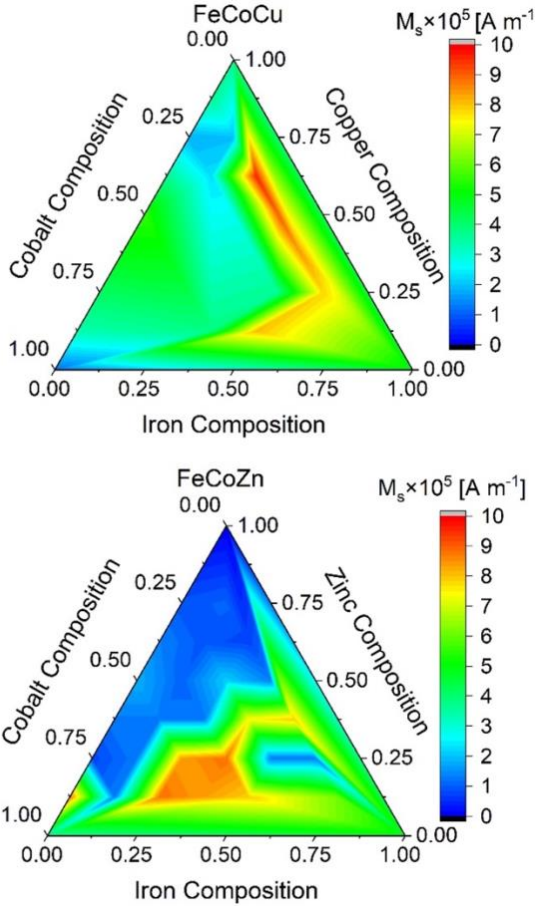
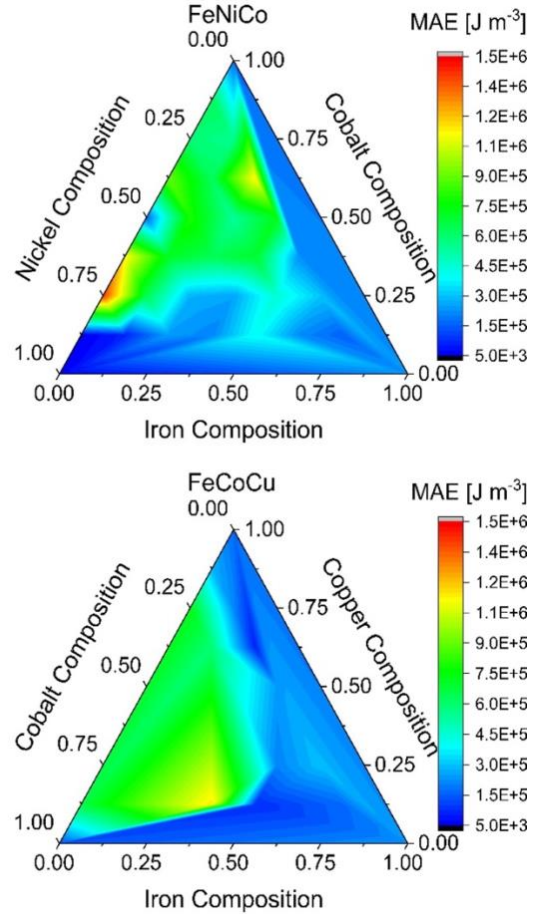

**Figure 3.** Calculated $M_s$ of FeCoCu (*top*) and FeCoZn (*bottom*). Fe and substituent compositions span from 0 to 16 and 0 to 8, respectively, with the stoichiometric ferrites represented at the vertices.

compositions hence show good promise for tuning $M_s$ through composition.

**3.3 Magnetic Anisotropy Energy.** Calculated MAE are shown in Figures 4 and S3 and tabulated in Table 1. MAE for the stoichiometric ferrites $Fe_{24}O_{32}$, $Mn_8Fe_{16}O_{32}$, $Co_8Fe_{16}O_{32}$, $Ni_8Fe_{16}O_{32}$, $Cu_8Fe_{16}O_{32}$, and $Zn_8Fe_{16}O_{32}$ are 2.2, 0.1, 1.7, 1.3, 1.3, and $0.6 \times 10^5$ J m$^{-3}$, respectively. We observe that similar to $M_s$, MAE is also dependent on composition; however, most compositions span within the same order of magnitude (Figures 4 and S3). For example, FeCu and FeMn only show variation in MAE when there is an almost equal number of substituent ions and Fe ions, possibly due to the resulting transformation of spinel structure (Table 1 and Figures S1 and S3 parts h and i). Exceptions are FeNiCo, FeCoZn and FeCoCu, presented in Figure 4 and S3b, which span three orders of magnitude. Comparing Figures S1 and S3, contrary to $M_s$, we observe that compositions that crystallize in inverse spinel result in higher MAE. The largest MAE value of $13.6 \times 10^5$ J m$^{-3}$ is found in FeNiCo (Figure 4a) in agreement with prior literature [55], and the lowest value is from $Zn_8Fe_{16}O_{32}$ of $0.06 \times 10^5$ J m$^{-3}$ (Figure S3b).

**Figure 4.** Calculated MAE (logarithmic scale) of FeNiCo (*top*) and FeCoCu (*bottom*). Fe and substituent compositions span from 0 to 16 and 0 to 8, respectively, with the stoichiometric ferrites represented at the vertices.

**4.0 Discussion.** Comparison of crystal structures, $M_s$, and MAE calculated in this work with experimental observation is provided in Tables 2 (specific compositions) and Section S1 (general compositions). Several compositions, e.g., $MnFe_2O_4$, $CuFe_2O_4$, and $CoFe_2O_4$ are in excellent agreement. Others, e.g., $Co_{4.8}Ni_{3.2}Fe_{16}O_{32}$, show notable differences. We have previously shown that these differences can be attributed to the simplicity of our models compared to real experimental

___

Table 2. Comparison of DFT calculated crystal structure, $M_s$, and MAE with experiment.

| Specific Composition | Crystal structure This work / expt | $M_s \times 10^5$ A m$^{-1}$ This work / expt | MAE $\times 10^5$ J m$^{-3}$ This work / expt |
|---|---|---|---|
| $Fe_{24}O_{32}$ | Inverse / Inverse [f] | 4.8 / 4.6 [f] | 2.2 / 0.1 [f] |
| $Fe_{16}Mn_8O_{32}$ | Normal / Normal [q] | 4.9 / 5.9 [q] | 0.1 / 0.2 [q] |
| $Fe_{16}Cu_8O_{32}$ | Inverse / Inverse [β] | 1.2 / 1.3 [β] | 1.3 / 1.4 [β] |
| $Fe_{16}Co_8O_{32}$ | Inverse / Inverse [§] | 3.7 / 3.5 [§] | 1.7 / 2.2 [£] |
| $Fe_{16}Ni_8O_{32}$ | Inverse / Inverse [€] | 3.6 / 2.0 [€] | 0.2 / 0.1 [f] |
| $Fe_{16}Zn_8O_{32}$ | Inverse / Normal [¥] | 0.04 / 0.09 [¥] | 0.6 / N/A |
| $Co_{1.6}Cu_{6.4}Fe_2O_{32}$ | Normal / N/A | 1.8 / 2.3 [Ω] | 7.4 / 3.6 [Ω] |
| $Co_{4.8}Ni_{3.2}Fe_{16}O_{32}$ | Inverse / N/A | 1.7 / 3.2 [Σ] | 9.0 / N/A |

[f]Ref. [17] ; [q]Ref. [58] ; [β]Ref. [59]; [Ω]Ref. [60]; [£]Ref. [61]; [¥]Ref. [4]; [§]Ref. [54]; [€]Ref. [62]; [Σ]Ref. [52]; N/A: not available

systems [56]. Specifically, our model systems are nearly pristine bulk structures, whereas experimental systems are particles with finite sizes, surfaces, and defects, as well as different ligands, crystallographic domains, etc. [1,10,56,57]. Further, experimental observation is an average value from a distribution of these properties, whereas the calculations presented herein are for individual structures. Unfortunately, it is not presently possible to model even one single nanoparticle with DFT, let alone a distribution for any one composition, and certainly not for a distribution of compositions. Hence, at present, a better use for these results is to learn how trends influence magnetic properties. To this end, Figure 5 shows calculated $M_s$ and MAE for the various doubly substituted non-stoichiometric ferrites.

Figure 5a shows that FeCoZn, FeCoCu, and FeNiCu can achieve relatively large ranges in $M_s$ from varying composition, whereas FeNiZn, FeNiCo, FeMnNi, and FeMnCo tend to have more uniform $M_s$. MAE for all compositions modeled in this work varies by ~ 2 orders of magnitude. To understand how these compositions could be used in practice, Figure 5b illustrates optimal combinations of $M_s$ and MAE for various applications. For example, our results suggest that FeMnCo and FeMnNi as well as some compositions of FeCoZn will be optimal for magnetic induction heating (evaluating these materials for toxicity for biomedical applications is a concern that is beyond the scope of this paper). MRI requires materials with high $M_s$ [9,11,63] and therefore FeCoZn, FeCoCu, and FeNiCu are promising. Permanent magnets require modest $M_s$ and high MAE [20,64] and hence FeMnNi and FeNiCo are promising. Ferrites with high $M_s$ and low MAE could potentially replace rare earth materials in antennas [10], and compositions such

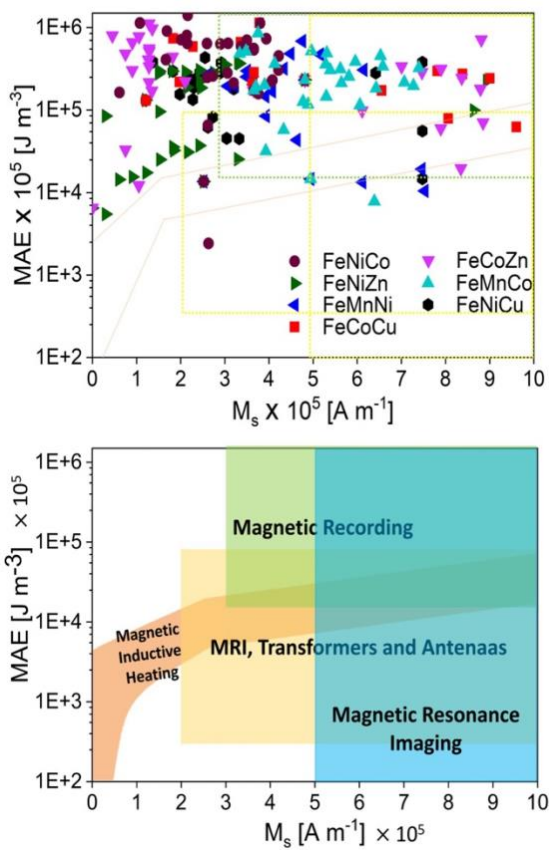

**Figure 5.** *Top:* Calculated $M_s$ and MAE (log scale). Dotted lines indicate the regions in the bottom graph. *Bottom:* Ranges of $M_s$ and MAE for various applications of magnetic materials.

___

as such as FeNiZn could achieve this. Transformers utilize magnetic induction and require minimal heat losses [65], i.e., high $M_s$ and low MAE. Based on Figure 5, FeCoZn and FeMnNi compositions could be promising.

**5.0 Conclusions.** In summary, structural and magnetic properties of non-stoichiometric bulk ferrites in the formula $M1_xM2_yFe_{3-x-y}O_4$ where M1 and M2 = Mn, Ni, Co, Cu, and/or Zn and $0 \leq x \leq 1$ and $y = 1 - x$ have been investigated using DFT. Through varying the composition, we found changes in crystal structure (from normal to inverse spinel), which resulted in variations in the magnetic saturation and magnetic anisotropy energy of $9.6 \times 10^5$ A m$^{-1}$ and $14.1 \times 10^5$ J m$^{-3}$, respectively. We found that magnetic properties are influenced by composition through their crystal structures, with normal spinel compositions resulting in higher $M_s$ and inverse spinel compositions resulting in higher MAE. Our results suggest that composition can be used to optimize magnetic properties for applications in heating, imaging, and recording. This is mainly achieved by varying $M_s$, as these applications are more sensitive to variation in $M_s$ than MAE (Figure 5). Moving forward, developing a strategy to achieve greater variation of MAE would lead to greater technological applicability. Our calculations suggest that doubly substituted non-stoichiometric ferrites based on Mn, Ni, Co, and Zn could achieve this. Comparison with available experimental data suggests DFT underpredicts $M_s$ and overpredicts MAE. As a major difference between experiments and our DFT simulations is that our simulations were performed on pristine bulk structures, whereas experiments were performed on nanoparticles comprising different crystallographic domains and surfaces with ligands and defects, these results suggest that a way to maximize control over magnetic properties in practice is to minimize these effects in order to have the greatest control over MAE while using composition to control $M_s$. This is a topic of ongoing work.

**Acknowledgments.** We thank Dr. Megan Hoover, Prof. Lindsay Shuller-Nickles, Prof. Steven Pellizzeri, Dr. Benjamin Fellows, and Dr. Zichun "Tony" Yan for helpful discussions. This work was partly supported as part of the Center for Programmable Energy Catalysis, an Energy Frontier Research Center funded by the U.S. Department of Energy, Office of Science, Basic Energy Sciences at the University of Minnesota under award #DE-SC0023464 (VRP & RBG: MAE calculations, magnetic property analysis, comparisons to experimental data, linking results to potential applications). We would also like to acknowledge support by Materials Assembly and Design Excellence in South Carolina (MADE in SC; VRP, JZ, PMM, AC: Model development, $M_s$ and crystal structure calculations), National Science Foundation award no. OIA-1655740 and Grants for Exploratory Academic Research (GEAR; OTM). We would also like to thank the support of National Science Foundation award no. CBET-2146591 (OTM). This work was supported in part by the Center for Integrated Nanotechnologies, an Office of Science User Facility operated for the U.S. Department of Energy (DOE) Office of Science (OTM).

**Keywords.** Structure function relationships, magnetic inductive heating, magnetic imaging, magnetic recording, spinel structures.

**Supporting Information.** Additional figures illustrating preferred crystal structure, calculated $M_s$, and calculated MAE; additional figures illustrating ranges of $M_s$ and MAE used in different applications; simulation input files; validations of the computational model and methods. This data can be obtained by emailing the corresponding author until the data is officially published in the peer reviewed literature (after which it will be freely available on the publisher's website).

_________________________________________________________________________________________________

---

______________________________________________________________________________

___________________________________________________________________________________________

___